\documentclass[aps,preprint]{revtex4}%
\usepackage{amsfonts}
\usepackage{amsmath}
\usepackage{amssymb}
\usepackage{graphicx}%
\begin{document}
\title{1D Cahn-Hilliard dynamics : coarsening and interrupted coarsening }
\author{Simon Villain-Guillot}
\affiliation{Laboratoire Onde et Mati\`{e}re d'Aquitaine, Universit\'{e} Bordeaux I }
\affiliation{351, cours de la Lib\'{e}ration 33405 Talence Cedex, France}
\affiliation{Email: simon.villain-guillot@u-bordeaux1.fr}

\begin{abstract}
Many systems exhibit a phase where the order parameter is spatially modulated.
These patterns can be the result of a frustration caused by the competition
between interaction forces with opposite effects.

In all models with local interactions, these ordered phases disappear in the
strong segregation regime (low temperature). It is expected however that these
phases should persist in the case of long range interactions, which can't be
correctly described by a Ginzburg-Landau type model with only a finite number
of spatial derivatives of the order parameter.

An alternative approach is to study the dynamics of the phase transition or
pattern formation. While, in the usual process of Ostwald ripening, succession
of doubling of the domain size leads to a total segregation, or
macro-segregation, C. Misbah and P. Politi have shown that long-range
interactions could cause an interruption of this coalescence process,
stabilizing a pattern which then remains in a micro-structured state or
super-crystal. We show that this is the case for a modified Cahn-Hilliard
dynamics due to Oono which includes a non local term and which is particularly
well suited to describe systems with a modulated phase.

\end{abstract}
\maketitle

\section{Introduction}

Many systems exhibit phases where the order parameter is spatially modulated
and forms a pattern \cite{SC}. These phases are the result of a frustration
caused by the competition between interaction forces with opposite effects.

For example, in a blend of polymers, the difference of interaction energies
between homo and hetero polymers generates locally a repulsion between
heteropolymers which leads to a macroscopic segregation. But for diblock
co-polymers which are built with two heteropolymers A and B which are attached
to each other by a chemical bond, such a macroscopic global phase separation
is prohibited. They form a disordered phase at high temperature (when the
entropic effects prevail), but below a critical temperature, whereas energetic
considerations should lead to segregation, this chemical binding prevents
separation between A and B heteropolymers over a long distance : the two
components A and B self-organized in patterns or domains of finite size
(mainly lamellar or hexagonal) in order to minimize nevertheless contacts
between heteropolymers en thus the energy of interaction. The relative density
in heteropolymers is thus spatially periodically modulated. This spontaneous
microstructuration could be helpfull to design a new generation of solar cells
based on organic semi-conductors\cite{amadeus}.

In all models with local interactions, these ordered phases disappear in the
strong segregation regime (low temperature). It is expected, however, that
these phases should persist in the case of long-range interactions, which
can't be correctly described by a Ginzburg-Landau type model with only a
finite number of spatial derivatives of an order parameter (which can be
defined in our preceding example from the relative density in the two
components A and B).

An alternative approach is to study the dynamics of phase transition. While,
in the usual process of Ostwald ripening, succession of coarsening events with
doubling of the domain size leads to a total segregation, or
macro-segregation, C. Misbah and P. Politi \cite{misbah} have shown that
long-range interactions could cause an interruption of this coalescence
process, stabilizing a pattern that remains consequently in a micro-structured
pattern or super-crystal.

We show here that this is the case for the equation of Oono\cite{brasil},
which is particularly well suited to describe the dynamics of systems with a
modulated phase.

\section{Dynamics of phase transitions}

\subsection{Time-Dependent Ginzburg Landau equation}

\subsubsection{Derivation of the model}

Different equations can be used to describe the dynamics of a phase transition
depending on, for example, if the order parameter is a scalar or a vector, and
whether it is conserved by the dynamics or not (for a review see
\cite{halperin,gunton}).

As at equilibrium, this order parameter must minimize a free energy, the
dynamics out of equilibrium must then involve deviation from this stable order
parameter value or function, just like in a simple mechanical system. The
simplest dynamics based on Ginzburg-Landau free energy for a scalar order
parameter is the TDGL (Time-Dependent Ginzburg Landau or model A in Hohenberg
and Halperin classification\cite{halperin}) which writes%
\begin{equation}
\frac{\partial u}{\partial t}\left(  \mathbf{r},t\right)  =\mathbf{-}%
\frac{\delta F_{GL}}{\delta u}=\mathbf{\nabla}^{2}u-\frac{\varepsilon}%
{2}u-2u^{3} \label{TDGL}%
\end{equation}

In this equation, $u\left(  \mathbf{r},t\right)  $ is a macroscopic order
parameter which is a coarse grained of a microscopic order parameter in a
small volume around the postition $\mathbf{r}$. And $\varepsilon$ is the
dimensionless control parameter, usually the reduce temperature $\varepsilon
=\frac{T-T_{c}}{T_{c}}$ where $T_{c}$ is the critical temperature of the phase
transition. This partial differential equation is invariant by the
transformations $u\rightarrow-u$ and $x_{i}\rightarrow-x_{i}+a_{i}$. $F_{GL}$
is the Ginzburg-Landau free energy local density or Lyapounov functional in
the context of dynamical systems :%
\[
F_{GL}=\frac{1}{2}\left(  (\mathbf{\nabla}u)^{2}+\frac{\varepsilon}{2}%
u^{2}+u^{4}\right)
\]

The non-local term $(\mathbf{\nabla}u)^{2}$ prevents discontinuity or
roughness of the order parameter and assigns energetic overcost to its
variations in proportion with their sharpness. When looking at the temporal
evolution of the free energy $\int F_{GL}(r,t)dr$~:%
\[
\frac{d}{dt}\int F_{GL}dr=\int\frac{\delta F_{GL}}{\delta u}.\frac{\partial
u}{\partial t}dr=\int\frac{\delta F_{GL}}{\delta u}.(-\frac{\delta F_{GL}%
}{\delta u})dr=-\int(\frac{\delta F_{GL}}{\delta u})^{2}dr<0
\]

One notices from equation \ref{TDGL} that the dynamics will induce a change of
$u\left(  \mathbf{r}\right)  $ as long as it hasn't reached a minimum of the
free energy density $F_{GL}$. If one looks for homogeneous states (where the
order parameter is independent of the spatial coordinates) to be stationary
states of this equation, they will be the extrema of the Landau potential
$V(u)=\frac{\varepsilon}{2}u^{2}+u^{4}$ which is plotted in Fig. 1 for the two
possible signs of the control parameter. For $\varepsilon>0$, the only
extremum is $u=0$, so there is only one homogenous solution, which is stable,
being a minimum of the Landau potential (which is a convex function as long as
$\varepsilon>0$). When $\varepsilon<0$, this potential is now concave in a
neighborhood of $u=0$, which is now a maximum and thus is now linearly
instable. Two other symmetric solutions $u=\pm\frac{\sqrt{-\varepsilon}}{2}$
have now appeared due to this pitchfork bifurcation. They are the new stable
homogeneous solutions and correspond to a minimum of the potential $V_{\min
}=-\varepsilon^{2}/32$.
%
%

\begin{figure}
[ptb]
\begin{center}
\includegraphics[
natheight=2.145600in,
natwidth=3.250000in,
height=2.1845in,
width=3.2949in
]%
{ImageGL.jpg}%
\caption{{\footnotesize Landau potential as a function of }$u${\small ,}
{\footnotesize the amplitude of the order parameter. We have plotted the
profil of this potential above and below the pitchfork bifurcation at
}$\varepsilon${\footnotesize =0.}\newline{\footnotesize For }$\varepsilon
>0${\footnotesize , the potential is a convex function and there is only one
minimum, }$u=0${\footnotesize .}\newline{\footnotesize For }$\varepsilon
<0${\footnotesize , the Landau potential is a concave function around }%
$u=0${\footnotesize , which is now a maximum ; two other solutions have now
appeared as miminum of the potential, symmetric one each other. }}%
\label{potlandau}%
\end{center}
\end{figure}
\pagebreak

\subsubsection{Linear stability analysis}

Linear stability analysis consists in computing the growth rate of small
fluctuations of a solution. When linearizing equation (\ref{TDGL}) around
$u=0$ (i.e. when neglecting the nonlinear term $u^{3}$) one gets%
\[
\frac{\partial u}{\partial t}\left(  \mathbf{r},t\right)  =\mathbf{-}%
\frac{\varepsilon}{2}u+\mathbf{\nabla}^{2}u
\]

Considering this equation in the Fourier space we can decompose $u$ in Fourier
series in the case of a finite size problem or Fourier transform in the
infinite case :%
\begin{equation}
u(\mathbf{r},t)=\sum_{\mathbf{q}}u_{\mathbf{q}}e^{i\mathbf{q\cdot r}+\sigma t}%
\end{equation}
where $u_{q}$ is the amplitude of the Fourier mode at $t=0$. For example, it
can be the thermal fluctuations proportional to $T$. This mode decomposition
enables to compute the $q$-dependence of the amplification factor $\sigma(q)$
(or growth rate or imaginary part of $k=q-i\sigma$)~:%
\begin{equation}
\sigma(\mathbf{q})=-(q^{2}+\frac{\varepsilon}{2})
\end{equation}

$\sigma(\mathbf{q})$ is negative for $\varepsilon>0$, and thus the homogeneous
solution $u=0$ is unstable with respect to fluctuations of the order
parameter. The whole band $0<q<\sqrt{(-\varepsilon/2)}$ is linearly unstable
as $\sigma(\mathbf{q})>0$ (see Fig. \ref{tdgl})%
\begin{figure}
[h]
\begin{center}
\includegraphics[
natheight=2.322900in,
natwidth=3.749800in,
height=2.3627in,
width=3.7974in
]%
{../../HDR/TDGL.jpg}%
\caption{{\small Amplification factor }$\sigma(q)${\small \ computed via
linear stabily analysis of the time-dependent Ginzburg-Landau equation (TDGL).
It is positive (growth of the modulations) for all the modes }$q<\sqrt
{\frac{-\varepsilon}{2}}${\small .}}%
\label{tdgl}%
\end{center}
\end{figure}
\pagebreak

\subsubsection{Symmetry breaking and conservation law}

The linear stability analysis enables to conclude that the most instable mode
is for $q=0$~: it is thus a long wave instability, which will give rise to
large homogeneous domains and imply spontaneous symmetry breaking.
This is the case, for example, in magnetic systems.

But if there is a conservation law, as for example a conservation of mass,
such an instantaneous symmetry breaking is prohibited : the matter, or the
different species diffuse with a finite characteristic time. Hillert
\cite{hillert}, Cahn and Hilliard \cite{CH} have proposed a model to describe
segregation in a binary mixture. This equation, later on denoted C-H for
Cahn-Hilliard, corresponds to model B in the Hohenberg and Halperin
classification\cite{halperin}. Cahn-Hilliard dynamics is the minimal equation
describing phase transition for a conserved scalar order parameter. As this
conservation law prevents global symmetry breaking, it will generate numerous
domains and interfaces separating them. This dynamic governs a whole class of
first order phase transition like the Fr\'{e}edericksz transition in liquid
crystals \cite{coullet}, segregation of granular media in a rotating
drum\cite{oyama}, or formation of ripple due to hydrodynamic oscillations
\cite{melo,stegner}.\pagebreak

\subsection{Model B or Cahn-Hilliard equation}

\subsubsection{Derivation of the model\bigskip}

Cahn-Hilliard dynamics is a modified diffusion equation for a scalar order
parameter $u$, which writes~:%

\begin{equation}
\frac{\partial u}{\partial t}\left(  \mathbf{r},t\right)  =\mathbf{\nabla}%
^{2}(\frac{\varepsilon}{2}u+2u^{3}-\mathbf{\nabla}^{2}u)=\mathbf{\nabla}%
^{2}(\frac{\delta F}{\delta u}) \label{CH}%
\end{equation}

In the original work of Cahn and Hilliard, $u\left(  \mathbf{r},t\right)  $
represents the concentration of one of the components of a binary alloy. But
it can also be the fluctuation of density of a fluid around its mean value, or
concentration of one chemical component of a binary mixture, or the height of
a copolymer layer\cite{copol}..

As in model A, this equation is invariant by the transformations
$u\rightarrow-u$ and $x_{i}\rightarrow-x_{i}+a_{i}$ and when looking at the
time evolution of the local quantity $F(t)$ , we still have~:%
\[
\frac{dF}{dt}=\frac{\delta F}{\delta\Phi u}.\frac{\partial u}{\partial
t}=\frac{\delta F}{\delta u}.\mathbf{\nabla}^{2}(\frac{\delta F}{\delta
u})=-(\mathbf{\nabla}\frac{\delta F}{\delta u})^{2}<0
\]

In order to derive a conservative dynamics, such that $\int\Phi(x,t)dx=cste$,
one can start from a detail balance \cite{langer} , or from a conservation
equation for the order parameter $\Phi$.%

\[
\frac{\partial u}{\partial t}=-\mathbf{\nabla}\cdot\mathbf{j}%
\]

where $\mathbf{j}$ is a matter current associated with $u$. This current is
related to the gradient of the chemical potential $\mu$ via the Hartley-Fick
law~: $\mathbf{j}=-\mathbf{\nabla}\mu$). And this chemical potential is itself
related to the functional derivative of the free energy $\mu=\frac{\delta
F}{\delta\Phi}$. This phenomenological approach enables to recover the C-H
equation(eq \ref{CH}).

If one looks globally at the quantity $\int u(x,t)dx=<u>$, the Cahn-Hilliard gives%

\[
\frac{d<u>}{dt}=\int\frac{\partial u}{\partial t}(x,t)dx=\int\mathbf{\nabla
}^{2}(\frac{\delta F}{\delta u}(x,t))dx=\left[  -(\mathbf{\nabla}\frac{\delta
F}{\delta u})\right]
\]

So, apart from boundary terms, the order parameter is indeed a conserved quantity.

\subsubsection{Linear stability analysis\bigskip}

Stationary states of the (C-H) are again the extrema of the Landau potential
$V(u)=\frac{\varepsilon}{2}u^{2}+u^{4}$. And after a quench, the system
undergoes a first order phase transition associated with the pitchfork
bifurcation from the $u=0$ solution to the symmetric solutions $u=\pm
\frac{\sqrt{-\varepsilon}}{2}$. But due to the conservation law, the dynamics
is different as Cahn and Hilliard have shown via the linear stability analysis
of equation (\ref{CH}) around $u=0$.%
\begin{equation}
\frac{\partial u}{\partial t}\left(  \mathbf{r},t\right)  =\mathbf{\nabla}%
^{2}\frac{\varepsilon}{2}u-\mathbf{\nabla}^{4}u \label{integrated}%
\end{equation}
one gets for the amplification factor in the Fourier space $\sigma(q)$ ~:%

\begin{equation}
\sigma(\mathbf{q})=-(q^{2}+\frac{\varepsilon}{2})q^{2} \label{lineaire}%
\end{equation}%
\begin{figure}
[ptb]
\begin{center}
\includegraphics[
natheight=2.293500in,
natwidth=3.666800in,
height=2.3333in,
width=3.7144in
]%
{../../figure3.bmp}%
\caption{{\small Amplification factor }$\sigma(q)${\small \ computed from the
linear stability analysis of Cahn and Hilliard equation}}%
\label{amplipli}%
\end{center}
\end{figure}

So, as $\sigma(\mathbf{q})$ is negative for $\varepsilon>0$, the $u=0$
solution is stable with respect to small fluctuations of the order parameter.
For negative $\varepsilon$, Fig \ref{amplipli} shows a band of instable
Fourier modes, as $\sigma(\mathbf{q})>0$ for $0<q<\sqrt{(-\varepsilon/2)}$.
Moreover, linear stability analysis of C-H predicts that the most instable
mode is not anymore for $q=0$ but for $q_{C-H}=\sqrt{-\varepsilon}/2$ (for
which $\sigma_{\max}=\frac{\varepsilon^{2}}{16}$). This wave number of maximum
amplification factor will dominate the first stage of the dynamics which is
called the spinodal decomposition; this explains in particular why the
homogeneous domains appear at length scales close to $L=\lambda_{C-H}%
/2=\pi/q_{C-H}$, half the wave length associated with the instability. For
longer times, interfaces separating each domain interact through Ostwald
ripening or coarsening, causing $<L>$ to change slowly toward higher values.

\section{Cahn-Hilliard equation}

\subsection{On the periodic solutions of Cahn-Hilliard equation}

When the equation is studied for a constant negative $\varepsilon$, via a
rescaling of $u$ (as $\sqrt{-\varepsilon}u$), position $\mathbf{r}$ (as
$\mathbf{r}/\sqrt{-\varepsilon}$) and time (as $t/{\left\vert \varepsilon
\right\vert }^{2}$), we observe that we could restrict the dynamics to the
case $\varepsilon=-1$. So later on, we will study the equation%

\begin{equation}
\frac{\partial u}{\partial t}\left(  \mathbf{r},t\right)  =\mathbf{\nabla}%
^{2}(-\frac{1}{2}u+2u^{3}-\mathbf{\nabla}^{2}u)
\end{equation}

In 1D, a family of stationary solution of this nonlinear dynamics is the
so-called interface-lattice solutions (or soliton-lattice), which writes~:
\begin{equation}
U_{k,\varepsilon}(x)=k\Delta\mathrm{Sn}(\frac{x}{\xi},k)\text{ with }%
\xi=\Delta^{-1}=\sqrt{2\left(  k^{2}+1\right)  } \label{amplitude}%
\end{equation}
where $\mathrm{Sn}(x,k)$ is the Jacobian elliptic function sine-amplitude, or
cnoidal mode. This family of solutions is parametrized by the Jacobian modulus
$k\in\left[  0,1\right]  $, or "segregation parameter". These solutions
describe periodic patterns of period%
\begin{equation}
\lambda=4K(k)\xi\text{, where }K(k)=\int_{0}^{\frac{\pi}{2}}\frac{\mathrm{d}%
t}{\sqrt{1-k^{2}\sin^{2}t}} \label{period}%
\end{equation}
is the complete Jacobian elliptic integral of the first kind. $K(k)$ together
with $k$, characterize the segregation, defined as the ratio between the size
of the homogeneous domains, $L=\lambda/2$, and the width of the interface
separating them, $2\xi$. The equation (\ref{period}) and the\ relation
$\xi=\Delta^{-1}$ enable to rewrite this family as :
\begin{equation}
U_{k,\lambda}(x)=\frac{4K(k)\cdot k}{\lambda}\mathrm{Sn}(\frac{4K(k)}{\lambda
}x,k). \label{ansatz}%
\end{equation}
and using equations (\ref{amplitude}) and (\ref{period}), we find that\ for a
stationary solution,\textbf{\ }$\lambda$\textbf{,\ }and $k$\ have to be
related one another through the following implicit equation (or the state
equation) :%
\begin{equation}
\lambda^{2}=2(1+k^{2})\left(  4K(k)\right)  ^{2}. \label{implicit}%
\end{equation}
Using equations (\ref{ansatz}) we can compute the free energy per unit length%
\[
\mathrm{F}_{GL}(k,\lambda)=
\]

\[
(\frac{4K}{\lambda})^{2}\left[  \frac{-\varepsilon}{4}(1-\frac{E}{K})+\left(
\frac{1+2k^{2}}{6}-\frac{E}{6K}(1+k^{2})\right)  (\frac{4K}{\lambda}%
)^{2}\right]
\]

where $E(k)$is the complete Jacobian elliptic integral of the second kind. The
absolute minimum for $\mathrm{F}_{GL}(k,\lambda)$ is for $k=1$ and
$\lambda=\infty$, i.e. for complete segregation with a single interface.

\subsection{Stationary States of the Cahn-Hilliard Dynamics}

The dynamics starts initially with $k=0$, for which $U(x)$ describes a
sinusoidal modulation of almost vanishing amplitude around the high
temperature homogenous stationary solution $u=0$
\begin{equation}
U_{k\rightarrow0,\varepsilon}(x)=k\sqrt{\frac{1}{2}}\sin(\sqrt{\frac{1}{2}}x)
\end{equation}

\[
=k\frac{2\pi}{\lambda_{C-H}}\sin(\frac{2\pi}{\lambda_{C-H}}x)=kq_{C-H}%
\sin(qx)
\]
The spinodal decomposition dynamics will saturate and reach a stationary state
which is a periodic pattern with a finite domain length (weak segregation
regime) for which $\lambda=\lambda_{C-H}$, and $k=k_{0}^{s}=\!0.687$ so as to
satisfy (\ref{implicit}), i.e $k$ is solution of the implicit equation :
\begin{equation}
2(1+k_{0}^{s2})K(k_{0}^{s})^{2}=-\frac{\varepsilon_{0}\lambda_{{\small C-H}%
}^{2}}{16}=\pi^{2}\text{ }.
\end{equation}
The amplitude of the modulation is then $k_{0}^{s}\Delta_{0}^{s}%
\!=\!0.400\sqrt{-\varepsilon_{0}}$, which is different from $u_{b}$.

Using linear stability analysis, Langer has shown that the stationary profile
thus obtained, $u_{0}(x)=U_{k_{0}^{s},\lambda_{C-H}}(x)$, is destroyed by
stochastic thermal fluctuations \cite{langer}. He has identified the most
instable mode as an \textquotedblright antiferro\textquotedblright\ mode,
leading to an infinite cascade of period doubling \cite{coarsening}. Disorder
of the pattern is also a cause of Ostwald ripening : if the periodicity of the
interface-lattice is broken, either when the distance between theses
interfaces or when the bulk value in the different domains become
non-constant, coarsening is triggered by diffusion of matter between
neighboring domains : big domains will then absorb smaller ones \cite{clerc}.

\subsection{Coarsening}

When considering the C-H equation \ref{CH} as a diffusion equation, P. Politi
and C. Misbah have shown that there should be coarsening as long as
\text{d}$\nu/$d$\lambda$ is positive, where $\nu$ is the amplitude of the
modulation and $\lambda$ its \cite{misbah}. As in Cahn-Hilliard dynamics
\[
\nu=k\Delta=k\sqrt{\frac{-\varepsilon}{2(k^{2}+1)}}\text{\ \ \ and\ \ }%
\ \lambda=4K(k)\xi=4K(k)\sqrt{\frac{2(k^{2}+1)}{-\varepsilon}}%
\]
are two growing functions of the parameter $k$, this diffusion coefficient
will always remain positive and coarsening will proceed until $\lambda
\rightarrow\infty$ (as in Fig. \ref{politi} left).%

\begin{figure}
[h]
\begin{center}
\includegraphics[
trim=0.000000in 0.300832in 0.000000in 0.000000in,
natheight=1.447700in,
natwidth=6.250000in,
height=1.1744in,
width=6.2768in
]%
{../../HDR/politi-misbah1.jpg}%
\caption{{\footnotesize Left : evolution of the amplitude of the modulation of
the stationart states as a function of the period, in the cases of a
Cahn-Hilliard dynamics. As d}$\nu/${\footnotesize d}$\lambda$%
{\footnotesize \ is always positive, the pattern will rippen until all the
interfaces disappear but one (note that as d}$\nu/${\footnotesize d}%
$\lambda\rightarrow0${\footnotesize , there is a slowing down of the
coarsening process). Right, a model where d}$\nu/${\footnotesize d}$\lambda
${\footnotesize \ changes sign : the coarsening will then be interrupted.}}%
\label{politi}%
\end{center}
\end{figure}
When looking at Figure (\ref{potlandau}), one can see that the bulk energy is
decreasing when the amplitude varies from $\nu=0$ to $\nu=\pm\frac
{\sqrt{-\varepsilon}}{2}$, that is, when the segregation increase. Meanwhile,
the interfacial energy is proportional to the period, we finally get that the
total energy decreases when the period of the stationary solutions gets longer
and longer. But for other dynamics (as in Fig. \ref{politi} right), d$\nu
/$d$\lambda$ can change of sign as we will see in the following : segregation
then remains partial. P. Politi et C. Misbah speak then of interrupted coarsening.

\section{Oono's model}

\subsection{Derivation of the model}

We would like to work out the period of modulated phase systems for which
there is a competition between two types of interactions: a short-range
interaction which tends to make the system more homogeneous together with a
long-range one, or a non-local one, which prefers proliferation of domain
walls. This competition results in a microphase separation with a preferred
mesoscopic length scale. These systems forming a super-crystal can be studied
using a modified Landau-Ginzburg approach, derived from Cahn-Hilliard
equation~and of practical use for numerical simulations \cite{brasil}:%
\begin{equation}
{\frac{\partial u}{\partial t}}=(\nabla^{2}\frac{\delta F_{GL}(u)}{\delta
u})-\beta^{2}u=\mathbf{\nabla}^{2}(\frac{-1}{2}u+2u^{3}-\mathbf{\nabla}%
^{2}u)-\left(  \frac{\beta}{4}\right)  ^{2}u. \label{oono}%
\end{equation}

The $-\beta^{2}u$ term models in the Cahn-Hilliard equation the long-range
interactions, which prevents the formation of macroscopic domains and favors
the modulation. We will see that the inclusion of such a term, following Oono,
enables to describe the behavior of modulated systems at $T$ much lower than
$T_{c}$. If we suppose, for example, that in a 3D problem, the long-range
interaction decreases like $\frac{1}{r}$, the full free energy density writes
\begin{equation}
F(u)=F_{GL}+F_{int}%
\end{equation}%
\[
=\frac{1}{2}(\nabla u(r))^{2}+\frac{-1}{4}u^{2}(r)+\frac{1}{2}u^{4}(r)+\int
u(r^{\prime})g(r^{\prime},r)u(r)\mathrm{d}r^{\prime}\text{ ,}%
\]

where $g(r^{\prime},r)=4\pi\frac{\left(  \frac{\beta}{4}\right)  ^{2}%
}{\left\vert r^{\prime}-r\right\vert }$ in D=3, or $\left\vert x^{\prime
}-x\right\vert $ in D=1$.$The long-range interaction $g(r^{\prime},r)$
corresponds to a repulsive interaction when $u(r^{\prime})$ and $u(r)$ are of
the same sign : thus it \ favors the formation of interphases. If we want to
study the dynamic of this phase separation, we use the Cahn-Hilliard
equation~:
\begin{equation}
{\frac{\partial u}{\partial t}}=\nabla_{r}^{2}\left(  \frac{\delta
F(u)}{\delta u}\right)
\end{equation}%
\[
=\nabla_{r}^{2}\left(  \frac{-1}{2}u+2u^{3}-\mathbf{\nabla}^{2}u+\int
u(r^{\prime})g(r^{\prime},r)\mathrm{d}r^{\prime}\right)  .
\]
If one recalls that $\frac{-1}{\left\vert r^{\prime}-r\right\vert }$ is the
Green's function associated with the Laplacian operator $\nabla_{r}^{2}$ in
3D, the preceding equation then transforms into
\begin{equation}
\nabla_{r}^{2}\left(  \int u(r^{\prime})g(r^{\prime},r)\mathrm{d}r^{\prime
}\right)  =\int u(r^{\prime})\nabla_{r}^{2}g(r^{\prime},r)\mathrm{d}r^{\prime}%
\end{equation}%
\[
=-\left(  \frac{\beta}{4}\right)  ^{2}\int u(r^{\prime})\delta(r^{\prime
},r)\mathrm{d}r^{\prime}=-\left(  \frac{\beta}{4}\right)  ^{2}u(r).
\]
which leads to equation (\ref{oono}). Note that, even with the new term added
by Oono to the usual Cahn-Hilliard dynamics, this equation remains in the
class of the conservative models, as it derives from a equation of
conservation. Note also that the free energy $F_{int}$ is infinite if $u(r)$
is of the same sign in a macroscopic domain.

\subsection{Linear stability analysis for Oono's model}

If we look at the linear stability analysis of the homogenous solution $u=0$,
we found almost the same results as in the original work of Cahn and Hilliard,
except that the amplification factor $\sigma(\mathbf{q})$~now write:%
\[
\sigma(\mathbf{q})=(\frac{1}{2}-\mathbf{q}^{2})\mathbf{q}^{2}-\left(
\frac{\beta}{4}\right)  ^{2}%
\]

This shows immediately that $u=0$ is linearly instable if $\beta<1$, with a
band of unstable Fourier modes $0.5\sqrt{1-\sqrt{1-\beta^{2}}}<q<0.5\sqrt
{1+\sqrt{1-\beta^{2}}}$ (for which $\sigma(\mathbf{q})>0$). The most unstable
mode is for $q_{C-H}=0.5$ like in the simplest Cahn-Hilliard model(\ref{CH}).
Therefore, during the initial stage of the dynamics, the spinodal
decomposition the homogeneous domains appear at length scales close to
$L=2\pi$, as in the usual Cahn Hilliard dynamics. But one sees that, contrary
to the simple Cahn-Hilliard case, the long wave length modulations are now
stable as $\sigma(\mathbf{q})<0$ for $q<0.5\sqrt{1-\sqrt{1-\beta^{2}}}$. This
explains qualitatively why, for any finite value of $\beta$, the dynamics will
end in a micro segregated regime, as it is observed numerically and as we will
discuss quantitatively below.

It has been noticed in different models \cite{kach} that, if the interaction
responsible of the modulation is local (i.e. described in the free energy by
local terms only, like $-(\mathbf{\nabla}u)^{2}$ in the Swift Hohenberg
model), then for low temperature or small $\beta$, the macrosegregated regime
(one unique interface) will be energetically favored compared to the
microphase separation.

However, in this model by Oono, because the interaction is long range (i.e.
non-local), no matter how small is $\beta$, there will always be a finite
region around $q=0$ where $\sigma(\mathbf{q})<0$. Indeed, $\sigma
(\mathbf{0})=-\left(  \frac{\beta}{4}\right)  ^{2}$. Consequently, a modulated
phase should always end the dynamics\cite{DA}.

\subsection{Direct minimization of the free energy}

For D=1, the contribution of the long-range interaction to the free energy per
unit length is \cite{gold}%
\[
\mathcal{F}_{int}=\frac{1}{\lambda}\int_{0}^{\lambda}F_{int}\mathrm{d}%
r=\frac{-\beta^{2}}{2\lambda}\int_{0}^{\frac{\lambda}{2}}\int_{0}%
^{\frac{\lambda}{2}}\Psi(r^{\prime})\left\vert r^{\prime}-r\right\vert
\Psi(r)\mathrm{d}r\mathrm{d}r^{\prime}\text{.}%
\]
When using as ansatz the family of interface-lattice solutions\ $U_{k,\lambda
}(x)$, we then obtain%
\begin{align*}
\mathcal{F}_{int}  &  =\frac{-\beta^{2}}{2\lambda}\int_{0}^{\frac{\lambda}{2}%
}\int_{0}^{\frac{\lambda}{2}}k^{2}(\frac{4K}{\lambda})^{2}\left\vert
r^{\prime}-r\right\vert \mathrm{Sn}(\frac{4K(k)}{\lambda}r,k)\mathrm{Sn}%
(\frac{4K(k)}{\lambda}r^{\prime},k)\mathrm{d}r\mathrm{d}r^{\prime}\\
&  =\frac{\pi}{K}\frac{-\beta^{2}}{8}\int_{0}^{2K}\int_{0}^{2K}k^{2}\left\vert
x^{\prime}-x\right\vert \mathrm{Sn}(x,k)\mathrm{Sn}(x^{\prime},k)\mathrm{d}%
x\mathrm{d}x^{\prime}.
\end{align*}
Thus, this contribution is independent of $\lambda\ $and the only minimization
is with respect to $k$. Consequently, the minimization with respect to
$\lambda$ concerns only $F_{GL}$ and enables to find $\lambda$ as a function
of $k$ : $\lambda(k)=8K\sqrt{\frac{1+k^{2}}{3}+\frac{k^{2}}{3(1-\frac{E}{K})}%
}$. And the minimization of the free energy $F_{GL}(k,\lambda(k))+F_{int}(k)$
is simply with respect to a single variable $k$, which can be done numerically
for different values of the interaction strength $\beta^{2}$.

Figure \ref{goldstonefig} presents $\lambda\left(  \beta^{2}\right)  $ which
scales like $\left(  \beta^{2}\right)  ^{1/3}$.%
\begin{figure}
[h]
\begin{center}
\includegraphics[
natheight=2.903200in,
natwidth=5.000400in,
height=2.3583in,
width=4.0413in
]%
{../../HDR/fig2.bmp}%
\caption{{\footnotesize Graph of the stable period }$\lambda\left(  \beta
^{2}\right)  ${\footnotesize \ computed by minimizing cthe free energy
}$F_{GL}(k,\lambda(k))+F_{int}(k,\beta^{2})${\footnotesize \ with respect to
}$k${\footnotesize . The result scales like }$\left(  \beta^{2}\right)
^{1/3}${\footnotesize .}}%
\label{goldstonefig}%
\end{center}
\end{figure}

\subsection{Stationary microsegregated patterns}

The family (\ref{ansatz}) is not anymore an exact stationary solution of the
dynamics (\ref{oono}) because of its last term. Nevertheless, it is a good
candidate for an approximate solution (especially in the case of small $\beta
$) and thus can be used as a tool for calculation using a solvability
condition or Fredholm's alternative.

Indeed, we can write deviation from a given periodic stationary profile of
period $\lambda$ as $u(x,t)=u_{0}(\phi(x,t))+\varepsilon u_{1}(\phi(x,t))+...
$ where $\varepsilon$ is a small parameter and $u_{0}$ is a periodic function
of the phase $\phi(x,t)$. For a steady state solution $\phi(x,t)=qx$ with
$q=2\pi/\lambda.\ $In the general case $\phi(x,t)=q(X,T)x $ where $X=\epsilon
x$ and $T=\epsilon^{2}t$ i.e. $q=\frac{\partial\phi}{\partial x}$ is now a
slowly varying function of $x$ and $t.$%

\begin{align*}
\frac{\partial u}{\partial t}  &  =\frac{\partial u}{\partial\phi}%
\frac{\partial\phi}{\partial t}=\frac{\partial u}{\partial\phi}\frac
{\partial\phi}{\partial T}\frac{dT}{dt}=\epsilon^{2}\frac{\partial\phi
}{\partial T}\frac{\partial u}{\partial\phi}\\
\frac{\partial u}{\partial x}  &  =\frac{\partial u}{\partial\phi}%
\frac{\partial\phi}{\partial x}=\frac{\partial u}{\partial\phi}(q+\frac
{\partial\phi}{\partial X}\frac{dX}{dx})=q\frac{\partial u}{\partial\phi
}+\epsilon\frac{\partial u}{\partial X}%
\end{align*}
If we denote $\Psi(X,T)=\epsilon\phi(x,t)$, then\ the local wave number is
$q(X,T)=\frac{\partial\phi}{\partial x}=\frac{\partial\Psi}{\partial X}$ and
\begin{align*}
\frac{\partial}{\partial t}  &  =\epsilon\partial_{T}\Psi\partial_{\phi}\\
\frac{\partial}{\partial x}  &  =q\partial_{\phi}+\epsilon\frac{\partial
q}{\partial X}\frac{\partial}{\partial q}=q\partial_{\phi}+\epsilon
\frac{\partial^{2}\Psi}{\partial X^{2}}\partial_{q}%
\end{align*}%
\begin{align*}
\frac{\partial^{2}}{\partial x^{2}}  &  =q\frac{\partial}{\partial\phi}\left(
q\partial_{\phi}+\epsilon\partial_{XX}^{2}\Psi\partial_{q}\right)
+\epsilon\partial_{XX}^{2}\partial_{q}\left(  q\partial_{\phi}+\epsilon
\partial_{XX}^{2}\Psi\partial_{q}\right) \\
\frac{\partial^{2}}{\partial x^{2}}  &  =q^{2}\partial_{\phi\phi}%
+\epsilon\partial_{XX}^{2}\Psi\partial_{\phi}+2\epsilon\partial_{XX}^{2}\Psi
q\partial_{q}\partial_{\phi}\\
\frac{\partial^{2}}{\partial x^{2}}  &  =q^{2}\partial_{\phi\phi}%
+\partial_{XX}^{2}\Psi\left(  1+2q\partial_{q}\right)  \partial_{\phi}%
\end{align*}
where we have kept only the first order terms in $\epsilon$.

If we consider a stationary profile $u_{0}$ which satisfies (zero order
equation):
\[
\ q^{2}\frac{\partial^{2}}{\partial\phi^{2}}\left(  \frac{-1}{2}u_{0}%
+2u_{0}^{_{^{3}}}-q^{2}\frac{\partial^{2}}{\partial\phi^{2}}u_{0}\right)
-\left(  \frac{\beta}{4}\right)  ^{2}u_{0}=0
\]%
\begin{equation}
\text{i.e. \ }\frac{\partial}{\partial\phi}\left(  \frac{-1}{2}u_{0}%
+2u_{0}^{_{^{3}}}-q^{2}\frac{\partial^{2}}{\partial\phi^{2}}u_{0}\right)
=\left(  \frac{\beta}{4}\right)  ^{2}w\text{ where }\partial_{\phi}%
w=q^{-2}u_{0} \label{primitiv}%
\end{equation}
Oono's equation (\ref{oono}) becomes then at order one in $\epsilon$%

\[
\epsilon\partial_{T}\Psi\partial_{\phi}u_{0}=\epsilon\mathcal{N}_{0}%
(u_{1})+\epsilon\mathcal{N}_{1}(u_{0})\text{ where}%
\]%
\begin{align*}
\mathcal{N}_{0}(u_{1})  &  =q^{2}\frac{\partial^{2}}{\partial\phi^{2}}%
(\frac{-1}{2}u_{1}+6u_{0}^{2}u_{1}-q^{2}\frac{\partial^{2}}{\partial\phi^{2}%
}u_{1})-\left(  \frac{\beta}{4}\right)  ^{2}u_{1}\\
&  =q^{2}\frac{\partial^{2}}{\partial\phi^{2}}\mathcal{L(}u_{1})-\left(
\frac{\beta}{4}\right)  ^{2}u_{1}\text{ and}%
\end{align*}%
\begin{align*}
\mathcal{N}_{1}(u_{0})  &  =\partial_{XX}^{2}\Psi\left(  1+2q\partial
_{q}\right)  \partial_{\phi}\left(  \frac{-1}{2}u_{0}+2u_{0}^{_{^{3}}}%
-q^{2}\frac{\partial^{2}}{\partial\phi^{2}}u_{0}\right) \\
&  -q^{2}\frac{\partial^{2}}{\partial\phi^{2}}\left(  \partial_{XX}^{2}%
\Psi\left(  1+2q\partial_{q}\right)  \partial_{\phi}u_{0}\right)
\end{align*}%
\[
=\left(  \frac{\beta}{4}\right)  ^{2}\partial_{XX}^{2}\Psi\left(
1+2q\partial_{q}\right)  w-q^{2}\frac{\partial^{2}}{\partial\phi^{2}}\left(
\partial_{XX}^{2}\Psi\left(  1+2q\partial_{q}\right)  \partial_{\phi}%
u_{0}\right)
\]
where we have used $\partial_{\phi}w=q^{-2}u_{0}$ and equation (\ref{primitiv}%
) to simplify $\mathcal{N}_{1}(u_{0})$. So Oono's equation (\ref{oono})
writes
\begin{equation}
\epsilon\partial_{T}\Psi\partial_{\phi}u_{0}-\left(  \frac{\beta}{4}\right)
^{2}\partial_{XX}^{2}\Psi\left(  1+2q\partial_{q}\right)  w \label{misbah1}%
\end{equation}%
\[
+q^{2}\partial_{XX}^{2}\Psi\frac{\partial^{2}}{\partial\phi^{2}}\left(
\left(  1+2q\partial_{q}\right)  \partial_{\phi}u_{0}\right)  =q^{2}%
\frac{\partial^{2}}{\partial\phi^{2}}\mathcal{L(}u_{1})-\left(  \frac{\beta
}{4}\right)  ^{2}u_{1}%
\]

\subsection{Stability of stationary microsegregated patterns}

A necessary condition for a solution to exist is that the left-hand side of
the system is orthogonal to the kernel of the adjoint operator $\mathcal{N}%
_{0}^{\dagger}=\left(  q^{2}\partial_{\phi\phi}\mathcal{L}-\left(  \frac
{\beta}{4}\right)  ^{2}Id\right)  ^{\dagger}$;\linebreak if $v\in$Ker$\left(
q^{2}\partial_{\phi\phi}\mathcal{L}-\left(  \frac{\beta}{4}\right)
^{2}Id\right)  ^{\dagger}$then the solvability condition (or Fredholm
alternative) writes:%
\[
<v|\partial_{T}\Psi\partial_{\phi}u_{0}-\mathcal{N}_{1}(u_{0})>=<v|\mathcal{N}%
_{0}(u_{1})>=0
\]
As for any $v$ we have
\begin{align*}
&  <v|q^{2}\frac{\partial^{2}}{\partial\phi^{2}}\left(  \frac{-1}{2}%
u_{1}+6u_{0}^{_{^{2}}}u_{1}-q^{2}\frac{\partial^{2}}{\partial\phi^{2}}%
u_{1}\right)  -\left(  \frac{\beta}{4}\right)  ^{2}u_{1}>\\
&  =<q^{2}\frac{\partial^{2}}{\partial\phi^{2}}v|\frac{-1}{2}+6u_{0}^{_{^{2}}%
}-q^{2}\frac{\partial^{2}}{\partial\phi^{2}})u_{1}>-\left(  \frac{\beta}%
{4}\right)  ^{2}<v|u_{1}>\\
&  =<q^{2}\left(  \frac{-1}{2}+6u_{0}^{_{^{2}}}-q^{2}\frac{\partial^{2}%
}{\partial\phi^{2}}\right)  \partial_{\phi\phi}v|u_{1}>-\left(  \frac{\beta
}{4}\right)  ^{2}<v|u_{1}>
\end{align*}
this adjoint operator writes :%
\[
\mathcal{N}_{0}^{\dagger}=\left(  q^{2}\partial_{\phi\phi}\mathcal{L}-\left(
\frac{\beta}{4}\right)  ^{2}Id\right)  ^{\dagger}=q^{2}\left(  \frac{-1}%
{2}+6u_{0}^{_{^{2}}}-q^{2}\frac{\partial^{2}}{\partial\phi^{2}}\right)
\partial_{\phi\phi}-\left(  \frac{\beta}{4}\right)  ^{2}%
\]

If $v\in$Ker$\mathcal{N}_{0}^{\dagger}$, we can define $\widetilde{u}$ such
that $q^{2}\partial_{\phi\phi}v=\widetilde{u}$ and which satisfies%
\begin{align}
q^{2}\frac{\partial^{2}}{\partial\phi^{2}}\left(  \frac{-1}{2}\widetilde
{u}+6u_{0}^{_{^{2}}}\widetilde{u}-q^{2}\frac{\partial^{2}}{\partial\phi^{2}%
}\widetilde{u}\right)   &  =q^{2}\left(  \frac{\beta}{4}\right)  ^{2}%
\partial_{\phi\phi}v=\left(  \frac{\beta}{4}\right)  ^{2}\widetilde
{u.}\label{astuce}\\
\text{So }\widetilde{u}\text{ is solution of }\frac{-1}{2}\widetilde{u}%
+6u_{0}^{_{^{2}}}\widetilde{u}-q^{2}\frac{\partial^{2}}{\partial\phi^{2}%
}\widetilde{u}  &  =\left(  \frac{\beta}{4}\right)  ^{2}v.\nonumber
\end{align}
Using equation (\ref{primitiv}), we thus find that $v$ defined by
\newline$q^{2}\partial_{\phi\phi}v($=$\widetilde{u})$=$\partial_{\phi}u_{0}$
is an element of \newline Ker$\left(  q^{2}\partial_{\phi\phi}\mathcal{L}%
-\left(  \frac{\beta}{4}\right)  ^{2}Id\right)  ^{\dagger}$.\newline As a
consequence, the diffusion equation writes
\[
\epsilon\partial_{T}\Psi=\frac{-q^{2}<v|\frac{\partial^{2}}{\partial\phi^{2}%
}\left(  \left(  1+2q\partial_{q}\right)  \partial_{\phi}u_{0}\right)
>+<v|\left(  \frac{\beta}{4}\right)  ^{2}\left(  1+2q\partial_{q}\right)
w>}{<v|\partial_{\phi}u_{0}>}\partial_{XX}^{2}\Psi
\]

As $q^{2}\partial_{\phi}w=u_{0}$ and $q^{2}\partial_{\phi\phi}v=\partial
_{\phi}u_{0}$ we get the equality%
\[
v=w.
\]%
\begin{align*}
\text{So }  &  <v|\partial_{\phi}u_{0}>=-<\partial_{\phi}v|u_{0}%
>=-<\partial_{\phi}w|u_{0}>\\
&  =-q^{-2}<u_{0}|u_{0}>
\end{align*}
and consequently equation (\ref{misbah1}) is a diffusion equation
\begin{align*}
\epsilon\partial_{T}\Psi &  =D\,\partial_{XX}^{2}\Psi\\
\epsilon\partial_{T}\Psi &  =q^{2}\frac{\partial_{q}<q\left(  \partial_{\phi
}u_{0}\right)  ^{2}>-\left(  \frac{\beta}{4}\right)  ^{2}\partial_{q}%
<q\,w^{2}>}{<u_{0}^{2}>}\partial_{XX}^{2}\Psi
\end{align*}

\section{Conclusion}

As long as the diffusion coefficient is negative (due to the $<\partial
_{k}u_{0}|\left(  \left(  1+2q\partial_{q}\right)  \partial_{\phi}%
u_{0}\right)  >=\partial_{q}<q\left(  \partial_{\phi}u_{0}\right)  ^{2}>$
term), the coarsening process goes on, in order to minimize interfacial
energy. But, due to its second part in $\beta^{2}$, the diffusion coefficient
will vanishe and thus the coarsening will be interrupted at a finite length scale.



\section*{Acknowledgment}

The authors would like to thank Dr. Chaouqi Misbah (LIPhy, Grenoble) for
fruitful discussions and an invitation in Grenoble where part of this work was done.





\end{document}